\documentclass[reprint,superscriptaddress,groupedaddress,a4paper,
nofootinbib,
nobibnotes,
 amsmath,amssymb,
pre,
floatfix,
showkeys,
]{revtex4-2}
\usepackage[dvipsnames]{xcolor}

\usepackage[utf8]{inputenc}
\usepackage{hyperref}
\usepackage{comment}
\usepackage{amsmath,amsfonts,bm,mathrsfs,dsfont,amssymb,xfrac,bigints,cancel,mathtools}
\usepackage{braket}
\usepackage{tikz}
\usetikzlibrary{patterns,arrows,shapes,arrows,intersections,decorations}

\renewcommand{\vec}[1]{\mathbf{#1}}

\newcommand{\grad}{\vec{\nabla}}

\renewcommand{\Re}{\operatorname{Re}}

\usepackage[oldvoltagedirection]{circuitikz}

\newcommand{\VD}{\ensuremath{ V_{\scriptscriptstyle D} }}
\newcommand{\ID}{\ensuremath{ I_{\scriptscriptstyle D} }}
\newcommand{\ZD}{\ensuremath{ Z_{\scriptscriptstyle D} }}
\newcommand{\VS}{\ensuremath{ V_{\scriptscriptstyle S} }}
\newcommand{\IS}{\ensuremath{ I_{\scriptscriptstyle S} }}

\DeclareMathOperator{\sech}{sech}

\begin{document}
\title{Feedback enhanced Dyakonov--Shur instability in Graphene-FET}
\author{Pedro Cosme}
\email{pedro.cosme.e.silva@tecnico.ulisboa.pt}
\affiliation{GoLP/Instituto de Plasmas e Fus\~ao Nuclear, Instituto Superior T\'ecnico, 1049-001 Lisboa, Portugal}
\author{Diogo Simões}
\affiliation{Instituto Superior T\'ecnico, 1049-001 Lisboa, Portugal}

\begin{abstract}

Graphene devices are known to have the potential to operate THz signals. In particular,
graphene field-effect transistors have been proposed as 
devices to host plasmonic instabilities in the THz realm; for instance, Dyakonov-Shur instability which relies upon
dc excitation. 
In this work, starting from a hydrodynamical description of the charge carriers, we extend the transmission line description of graphene field-effect transistors to a scheme with a positive feedback loop, also considering the effects of delay, which leads to the transcendental transfer function with terms of the form $e^{a s}\sech^k(s)/s$.
Applying the conditions for the excitation of Dyakonov--Shur instability, we report an enhanced voltage gain in the linear regime that is corroborated by our simulations of the nonlinear hydrodynamic model for the charge carriers. This translates to both greater saturation amplitude -- often up to 50\% increase -- and fastest growth rate of the self-oscillations. Thus, we bring forth a prospective concept for the realization of a THz oscillator suitable for future plasmonic circuitry.
\end{abstract}

\date{\today}
\maketitle

\section{Introduction}
Present day technology -- and thus society itself -- relies heavily on high frequency communication and nanoscale devices; yet, despite all the advances in this subject the THz range of the spectrum, with the potential to bridge the realms of electronics and photonics, still proves to be a challenge. Thus, new and efficient ways to generate, detect, or manipulate electromagnetic radiation in this range are ever sought after.

In this context, several designs of graphene nano-devices have been proposed \cite{Cosme2020,Cosme2021,Aizin2016Current-drivenNanostructures,Ryzhii2005,Ryzhii2008,Ryzhii2022DiscoveryPlasmons,Otsuji2014a,Safari2020,Yadav2018a} where an instability triggers the current to self-oscillate. Amongst the various mechanisms, Dyakonov--Shur (DS) instability \cite{Dyakonov1993,Crowne2000DyakonovShurTransistor}, driven
by the large asymmetry of impedance at the contacts, has been frequently brought forth as a way to induce self-oscillations from dc current. However, such instability has a modest amplitude gain, to circumvent that we study the use of a feedback loop as a way to improve the gain
\cite{Cosme2022Closed-LoopEmission} as well as the growth rate. 

In this work, we consider the effect of positive feedback on a graphene field-effect transistor (gFET) under the conditions for DS instability. For the analytical characterization,
we resort to a transmission line description \cite{Aizin2012TransmissionStates,Aizin2019PlasmonsApproach} that can be obtained from the hydrodynamic description of charge carriers. Alongside fully nonlinear simulations performed by \textsc{tethys} simulation tool \cite{Cosme2023}


A growing number of both theoretical \cite{Lucas2018,Muller2009,Fritz2023HydrodynamicTransport,Narozhny2019ElectronicGraphene,Kiselev2020NonlocalFluids} and experimental works \cite{Monch2022RatchetTransport,Ku2020a,Kumar2022ImagingResistance,Denisov2022SpinFluid,Vool2021ImagingWTe2,Denisov2022SpinFluid,Krebs2023ImagingFluids} have shown that the hydrodynamic regime of carriers in graphene and similar 2D materials is achievable even around room temperature. Therefore, the mean-field description of the carriers number density $n$ and momentum density $\vec{p}=nm^\star\vec{v}$ can be written as a set of conservation equation in the form \cite{Cosme2021,Cosme2020}:
\begin{subequations}
\begin{equation}
 \frac{\partial n}{\partial t} + \grad \cdot \frac{\mathbf{p}}{m^\star} = 0\text{ and}
\end{equation}
 \begin{equation}
 \frac{\partial\mathbf{p}}{\partial t} + \grad\cdot \left( \frac{\mathbf{p}\otimes\mathbf{p}}{nm^\star} + \frac{v^2_F}{3}n^{3/2} + \frac{c^2}{2} n^2 \right)=0,
\end{equation}\label{eq:fluidmodel}%
\end{subequations}%
where $v_F=10^6{\rm\,ms^{-1}}$ is the Fermi velocity and $c=$ the characteristic velocity of the plasma oscillations in the gradual-channel approximation\cite{S.M.Sze2007c}; moreover, due to the linear nature of the bands in graphene the mass is defined via the Drude weight $m^\star = \hbar\sqrt{\pi n}/v_F$.
This system undergoes DS instability when the boundary conditions 
\begin{equation}
    n(t,0)=n_0\text{ and }n(L,t)v(L,t)=n_0v_0,
\end{equation}
corresponding to a ac short ac open \cite{Dyakonov1993,Crowne2000DyakonovShurTransistor}, are applied.
Assuming uniformity along the direction transverse to the charge flow and in the gradual-channel approximation, we can convert to the local description to the macroscopic quantities of interest of current and voltage at source $(x=0)$ and drain $(x=L)$ by
$I(x)=-eW v(x)n(x)$ and $V(x)=-en(x)/C_g$, with $W$ the gFET width and $C_g$ its gate capacitance per unit area.

\section{Transfer matrix description of the Graphene transistor}\label{sec:transfer}
\subsection{Open-Loop Transfer Function}

\begin{figure}[!ht]
    \centering
    \includegraphics[width=.85\linewidth]{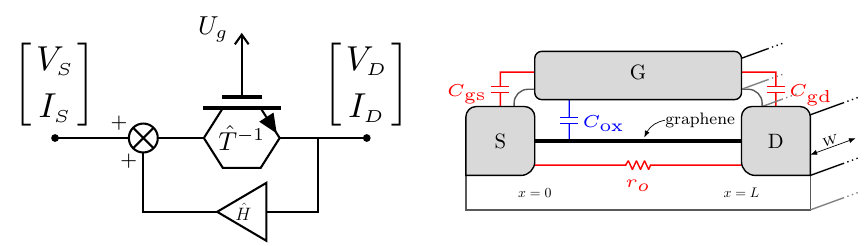}
    \caption{Left panel: Schematic rendition of the positive feedback loop. Right panel: Simplified model of the gFET considered, highlighting the gate-channel capacitance $C_{\rm ox}$ as well as the main parasitic impedances  represented.}
    \label{fig:feedback_diagram}
\end{figure}
 
From the hydrodynamical description given in \eqref{eq:fluidmodel} one can derive a transfer matrix description (see details in Refs.~\cite{Aizin2012TransmissionStates,Aizin2019PlasmonsApproach}) in the form 
\begin{equation}
\begin{bmatrix}
\VS\\
\IS
\end{bmatrix}\! = \hat{T}\begin{bmatrix}
\VD\\
\ID
\end{bmatrix}
\text{with }
\hat{T} = e^{\theta s} \begin{bmatrix} 
\cosh s &  Z_0\sinh s\\
 Z_0^{-1} \sinh s &   \cosh s
\end{bmatrix}\!,
\end{equation}
akin to a transmission line, with $\theta=v_0/v_p$ the ratio between the drift velocity and the plasmons velocity, the characteristic impedance $Z_0^{-1}=C_{g}Wv_p$, and $s= i\frac{L v_p \omega}{v_0^2-v_p^2} $ the normalized complex frequency.  
Thus, we can retrieve the voltage gain 
\begin{equation}
    \frac{\VD}{\VS}=e^{-\theta s}\left(\cosh s + \frac{Z_0}{\ZD(s)}\sinh  s\right)^{-1}=g(s)\label{eq:ganho_aberto}
\end{equation}
making use of the drain impedance $\ZD=\VD/\ID$, which for an ideal ac open circuit, as required for Dyakonov--Shur, $\ZD=\infty$. For non-ideal impedance matching, one can consider a typical a common gate configuration where $\ZD^{-1}\approx s C_{\rm g d}+r_o^{-1}$ and append such poles as $|\VD/\VS|\approx \left|\frac{
e^{\theta s} \sech s}{1-Z_0/\ZD}\right|$ for $|Z_0/\ZD|\ll 1$, thus, for the sake of simplicity, in the remainder of this work we assume $\ZD \to \infty$.

As we can see, all the poles of $g(s)$ are located along the imaginary axis, rendering the system marginally stable and in fact any perturbation leads to its instability. Moreover, note that, the unit step response of \eqref{eq:ganho_aberto} is a square wave \cite{1954Tables1} as
\begin{equation}
  \frac{\sech s}{s} \xrightarrow{\mathcal{L}^{-1}} \begin{cases}
    2,  & 1<t<3\pmod4\\
    0, & \text{otherwhise}
  \end{cases},\label{eq:square_wave}
\end{equation}
a trait that is clear in our simulations during the linear regime.

\subsection{Closed-Loop Transfer Function}

Let us now assume that a feedback loop is imposed on our system, as represented in Fig.\ref{fig:feedback_diagram}. In general, we have 
\begin{equation}
\begin{bmatrix}
\VS\\
\IS
\end{bmatrix}=
(\hat{T}-\hat{H}) \begin{bmatrix}
\VD\\
\ID
\end{bmatrix}
\end{equation}
where the matrix $\hat{H}$ encodes the proportion of voltage and/or current that is looped. solving for $\VD$ leads to the closed loop gain: 
\begin{equation}
\frac{\VD}{\VS}\equiv G(s)=\frac{1}{e^{\theta s}\cosh s-\beta e^{-\tau s}},
\end{equation}
with $\beta<1$ the voltage fraction and $\tau$ a delay factor. 
Since for any experimental implementation of feedback, the source and drain of the gFET will have to be connected by conductive material, its effect
can not be instantaneous, but rather to take a given time delay $\tau$ to occur, such delay is estimated to be of the order of the travel time of a 
a plasmon, i.e. $\tau \gtrsim L/v_p$.

\begin{figure}[!ht]
    \centering
    \includegraphics[width=.85\linewidth]{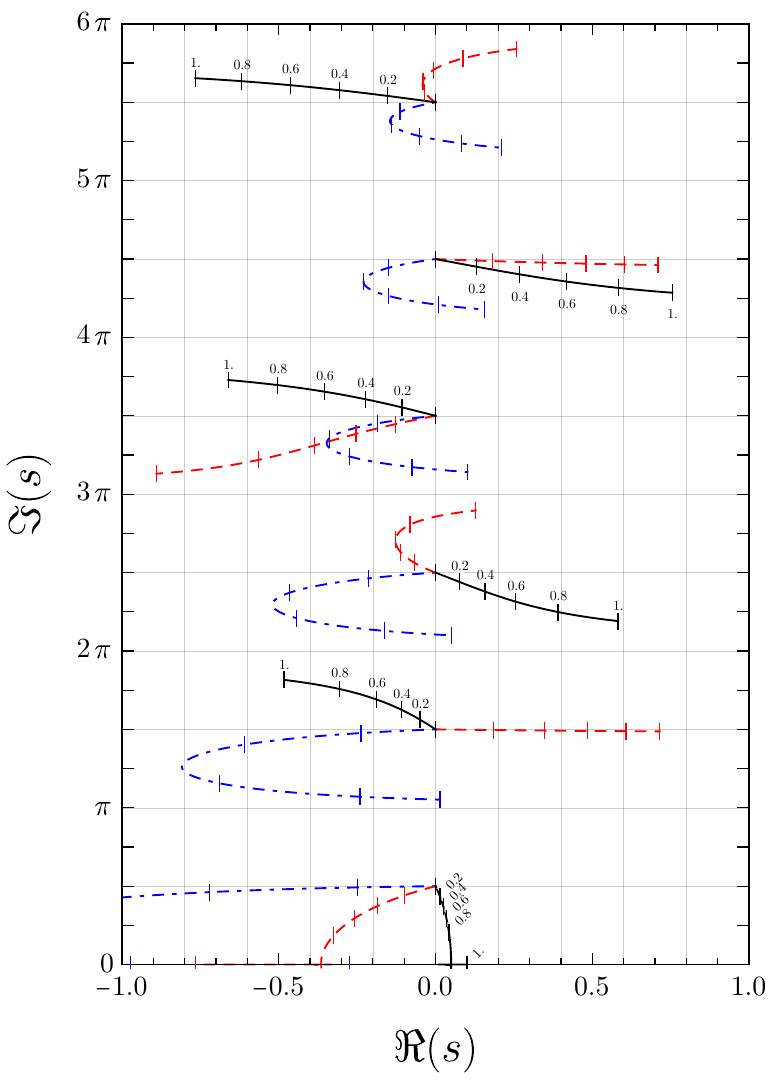}
    \caption{Location of the first six poles of the closed-loop transfer function $G(s)$ on the $s$-plane as function of feedback ratio, with $\theta\equiv v_0/v_p=1/20$. 
    Ticks along the curves indicate the ratio $\beta$ in $0.2$ increments, and the values of delay are $\tau=0$ (solid black lines); $\tau=0.4$ (red dashed lines); and $\tau=1$ (blue dot-dashed lines).}
    \label{fig:poles}
\end{figure}

\begin{figure}[ht!]
    \centering
    \includegraphics[width=\linewidth]{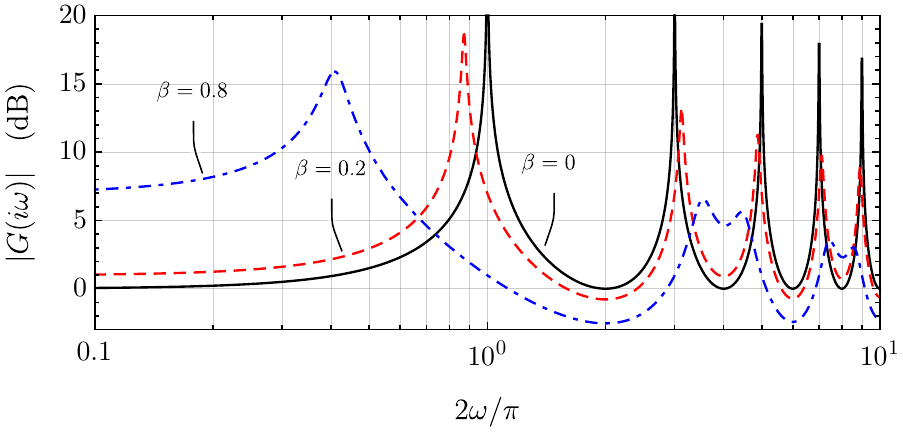}
    \includegraphics[width=\linewidth]{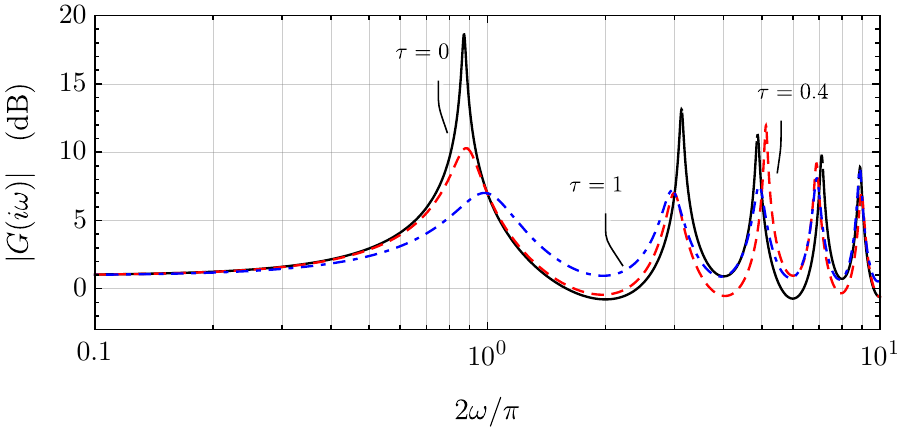}
    \caption{Bode plot for the closed-loop transfer function $G(s=i\omega)$ with $\theta\equiv v_0/v_p=1/20$. Top panel: Instantaneous feedback comparing the effect of feedback ratio. Bottom panel: Effect of delay time $\tau$ for fixed ratio of $\beta=0.2$ .}
    \label{fig:bodeplots}
\end{figure}

The feedback drastically modifies the structure of the transfer function, and the location of the poles start to wander over the complex plane, as can be seen in Fig.\ref{fig:poles}, for moderate values of delay the odd poles drift towards the unstable right-half plane while the even ones recede becoming decaying modes -- a feature that might be exploited for upconversion. 
Furthermore, as expected the feedback pushes the overall gain up, as is made clear by the Bode plots on Fig.\ref{fig:bodeplots}; the delay, however, appears to curb the self-amplification. 

Regarding the unit step response of the looped system, we expand the transfer function in powers of the feedback amplification, which leads to
\begin{equation}
    \frac{G(s)}{s}= \frac{e^{-\theta s} \sech s}{s} + \beta\frac{e^{-s(2\theta + \tau)} \sech^2 s}{s} + \mathcal{O}(\beta^2).\label{eq:beta_expand}
\end{equation}
So, while the first term yields the usual square wave, the subsequent terms in the series return unbounded functions (cf. appendix \ref{app:laplace} and Refs.~\cite{1954Tables1,Gardner1942TransientsI}), as 
$\mathcal{L}^{-1}[\sech^k(s)/s]\asymp t^{k-1}$, this leads to a faster growth rate for the instability but evidently cannot represent physical modes of the solutions for $t\to\infty$.
Nonetheless, one must bear in mind that insofar we have discussed but linear analysis of a nonlinear system, and the transient growing terms of \eqref{eq:beta_expand} saturate at a certain point.

\section{Nonlinear Simulations}

Thus far, the transfer function analysis provided crucial insight for the stability and early stages of the gFET operation under DS instability. However, aiming for  continuous the THz generation, the time evolution for later times is critical. To surpass the limitations of the linear analysis, the hydrodynamic system \eqref{eq:fluidmodel} needs to be numerically simulated so that the full nonlinear effects could manifest.   
We resorted to \textsc{Tethys} computer program \cite{Cosme2023}, a finite-volume solver designed for the simulation of graphene electrohydrodynamics. 

\begin{figure}[ht!]
    \centering
    \includegraphics[width=.85\linewidth]{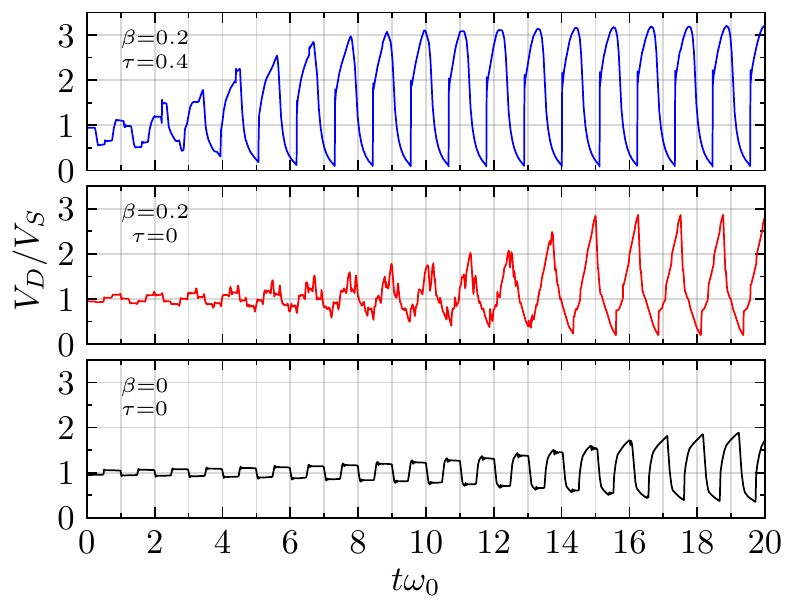}
    \caption{Nonlinear oscillations due to DS instability with delayed feedback. Direct simulation of the fluid model \eqref{eq:fluidmodel} with $S/v_0=20$. Note the growth of the saturation amplitude and the quicker onset of the saturated regime.}
    \label{fig:voltage_time}
\end{figure}

Under the DS conditions, we obtain the typical results (cf. \cite{Cosme2020,Satou2016,Crabb2021HydrodynamicTransistors,Daneshmandian2020,Cheremisin2002}) for the voltage, which are presented on Fig.\ref{fig:voltage_time} and where is clear the positive impact of the feedback albeit the modest $\beta$ ratio.

Even in the absence of feedback, one can distinguish three regimes, viz. linear evolution, transient time, and saturation. 
For the first few periods the system roughly follows a square wave, as predicted by \eqref{eq:square_wave}, before reaching saturation, which for the $\beta=0$ is dictated by Rankine--Hugoniot jump considerations that limit the maximum amplitude -- and thus the power diverted to the ac response.       
To the contrary, closing the feedback loop leads to a  significant increase in the saturation amplitude and to a fastest saturation itself. However, the waveform is distorted into a new shape due to the interplay of the higher order modes in the expansion \eqref{eq:beta_expand}.  

\section{Conclusion}

Since the discovery of graphene and its first applications to integrated circuit devices the number of uses and designs is ever-growing, boosted also by the discovery and development of new 2D materials. 

In that context, our study -- of a graphene transistor under a feedback loop -- paves the way for future self-driven oscillators in the THz range. We report that the effect of the closed loop unto Dyakonov Shur instability mitigates the usual technical problems, since it enhances both the growth rate and the amplitude of oscillations. The former is essential to overcome the eventual damping mechanisms ever present in real devices; while the latter improves the power and resilience of the signal. 

Furthermore, resorting to the transfer matrix description, we characterized the delayed closed loop system. 
Firstly, the trajectory of the poles on the complex plane show that the feedback leads to further destabilization of the odd spectral modes -- moving rightwards -- in detriment of the even ones -- that move to the left half plane. Such effect explain the reported third harmonic intensification \cite{Cosme2022Closed-LoopEmission} and can be further tuned to place this third harmonic on the goal frequency. 
Secondly, expanding the transfer function in powers of the feedback ratio evinces the structure of the unit step response as a sum of square waves whose amplitude grows with a power law, since the inverse Laplace transform of $G(s)/s$ scales as $t^2 $ up to second order in the feedback ratio $\beta$.
Nonetheless, as expected, the delay must be tailored to avoid destructive interference between the waves imposed by the loop and the source terminal and the counter-propagating ones inside the channel. 

The simulation of the nonlinear stage complemented our linear analysis and proved the increase of the saturation amplitude, e.g. from a voltage ratio $\VD/\VS\approx 2$ to $\VD/\VS\approx 3$,  by applying a 20\% feedback. However, this saturation stage and its limit cycle is still open for further inquiries, namely: which Liénard equation would accurately describe it, and how does the limit cycle vary with the system parameters.

Lastly, throughout this work, by using the classic transmission line model, we implicitly assumed galilean invariance of the plasmons. This constitutes a good approximation, since for most systems the drift velocities are well below the Fermi velocity. Yet, in future works the deviation from this assumption could be an interesting line of research to pursue. 

\begin{acknowledgments}
Lastly, one of the authors (PC) gratefully acknowledges the financial support of FCT -- Fundação para a Ciência e a Tecnologia (FCT--Portugal) through grant PB/DB/150415/2019
\end{acknowledgments}

\bibliographystyle{apsrev4-2}
\bibliography{references}

\appendix

\section{Inverse Laplace transform}\label{app:laplace}

The computation of the unit step response for the closed-loop system resorst to inverse Laplace transforms of powers of the form $\mathcal{L}^{-1}[\sech^k(s)/s]$. Here, in Tab.\ref{tab:tabela_laplace}, we provide closed form expressions for the first few terms, that attest the polynomial growth for the envelope of the base square wave.  
\begingroup
\renewcommand{\arraystretch}{2} 
\begin{table}[!ht]
    \centering
        \caption{First six inverse Laplace transform of $\sech^k(s)/s$.}
    \begin{ruledtabular}
    \begin{tabular}{cc}
         $\displaystyle F(s)$ & $\displaystyle f(t)=\mathcal{L}^{-1}[F(s)]$ \\[.5ex]
         \hline\\[-1em]
          $\displaystyle\frac{\sech s}{s}$ & $\displaystyle  1-(-1)^{\left\lfloor \frac{t+1}{2}\right\rfloor}$\\[3ex]
          $\displaystyle\frac{\sech^2 s}{s}$ & $\displaystyle 1 + (-1)^{\left\lfloor \frac{t}{2}\right\rfloor}\left(\! -2\left\lfloor \frac{t}{2}\right\rfloor -1\right)$\\[3ex]
          $\displaystyle\frac{\sech^3 s}{s}$ &  $\displaystyle 1+(-1)^{\left\lfloor \frac{t+1}{2}\right\rfloor}\left(2\left\lfloor \frac{t+1}{2}\right\rfloor ^2-1\right)  $\\[3ex]
           $\displaystyle\frac{\sech^4 s}{s}$ & $\displaystyle 1+(-1)^{\left\lfloor \frac{t}{2}\right\rfloor}\left(\frac{4}{3} \left\lfloor \frac{t}{2}\right\rfloor ^3+2 \left\lfloor \frac{t}{2}\right\rfloor ^2-\frac{4}{3} \left\lfloor \frac{t}{2}\right\rfloor -1\right)$\\[3ex]
    $\displaystyle\frac{\sech^5 s}{s}$ & $\displaystyle 1+(-1)^{\left\lfloor\! \frac{t+1}{2}\!\right\rfloor}\!\left(\!-\frac{2}{3} \left\lfloor \frac{t+1}{2}\right\rfloor ^4+\frac{8}{3} \left\lfloor \frac{t+1}{2}\right\rfloor ^2-1\right)$\\[3ex]
    $\displaystyle\frac{\sech^6 s}{s}$ & $\displaystyle1+(-1)^{\left\lfloor \frac{t}{2}\right\rfloor}\bigg(\!\!-\frac{4}{15} \left\lfloor \frac{t}{2}\right\rfloor ^5-\frac{2}{3} \left\lfloor \frac{t}{2}\right\rfloor^4+\frac{4}{3} \left\lfloor \frac{t}{2}\right\rfloor ^3+$\\[1.5ex]&\hfill$\displaystyle+\frac{8}{3}
   \left\lfloor \frac{t}{2}\right\rfloor ^2-\frac{16}{15} \left\lfloor \frac{t}{2}\right\rfloor -1\bigg)$
    \end{tabular}
    \end{ruledtabular}
    \label{tab:tabela_laplace}
\end{table}
\endgroup

\end{document}